\documentclass[letter,twocolumn]{jpsj2} 

\usepackage{times}
\usepackage{graphicx}

\title{Multipole State of Heavy Lanthanide Filled Skutterudites}

\author{Takashi {\sc Hotta}}

\inst{Advanced Science Research Center,
Japan Atomic Energy Agency, Tokai, Ibaraki 319-1195}

\recdate{\today}

\abst{
We discuss multipole properties of filled skutterudites containing
heavy lanthanide Ln from a microscopic viewpoint
on the basis of a seven-orbital Anderson model.
For Ln=Gd, in contrast to naive expectation, quadrupole moments
remain in addition to main dipole ones.
For Ln=Ho, we find an exotic state governed by octupole moment.
For Ln=Tb and Tm, no significant multipole moments appear
at low temperatures, while for Ln=Dy, Er, and Yb,
dipole and higher-order multipoles are dominant.
We briefly discuss possible relevance of these multipole states
with actual materials.
}

\kword{multipole, filled skutterudites, heavy lanthanide,
numerical renormalization group method}

\begin{document}
\maketitle


Recently, filled skutterudite compounds LnT$_4$X$_{12}$
with lanthanide Ln, transition metal atom T, and pnictogen X
have attracted much attention in the research field
of condensed matter physics due to large varieties of
exotic phenomena concerning magnetism and superconductivity.
\cite{Sato,Aoki1}
Since these compounds crystallize in the cubic structure
with high symmetry of $T_{\rm h}$ point group,\cite{Takegahara}
$f$-orbital degeneracy remains in general.
Due to the strong spin-orbit coupling of $f$ electrons,
spin-orbital complex degrees of freedom, i.e., {\it multipoles},
become active in filled skutterudite compounds,
which is a characteristic issue of this material group.

In fact, recent experiments in cooperation with
phenomenological theory have revealed that
multipoles appear ubiquitously in filled skutterudites.
For instance, a rich phase diagram of PrOs$_4$Sb$_{12}$ including
field-induced quadrupole order has been unveiled.
\cite{Aoki2,Tayama,Shiina}
In NdFe$_4$P$_{12}$, a significant role of quadrupole at low temperatures
has been suggested from the measurement of elastic constant.\cite{Nakanishi}
In SmRu$_4$P$_{12}$, a possibility of octupole order has been
proposed from experimental results.
\cite{Yoshizawa,Hachitani,Masaki1,Masaki2}
Recently, antiferro $\Gamma_1$-type order \cite{Kuramoto}
has been actively discussed experimentally and theoretically
for PrRu$_4$P$_{12}$\cite{Takimoto,Iwasa}
and PrFe$_4$P$_{12}$.\cite{Kiss,Sakai,Kikuchi}

In such current circumstances, it is one of challenging issues
to clarify microscopic aspects of multipole degrees of freedom
in $f$-electron systems.
The present author and collaborators have developed
a microscopic theory for multipole-related phenomena
on the basis of a $j$-$j$ coupling scheme.\cite{Hotta1}
In particular, possible multipole states for light lanthanide
filled skutterudites have been discussed by analyzing
multipole susceptibility in detail.
\cite{Hotta2,Hotta3,Hotta4,Hotta5,Hotta6,Hotta7,Hotta8}

However, multipole phenomena in heavy lanthanide systems
have not been understood satisfactorily, mainly due to the
difficulty in the theoretical treatment of multipoles
in lanthanides with large total angular momentum.
Experimentally, due to the development of crystal growth
technique under high pressure, heavy lanthanide filled skutterudite
compounds have been synthesized \cite{Sekine,Sekine2,Shirotani}
and research activities to clarify multipoles in such materials
will be advanced.
It is required to make more effort in the theoretical side
to develop a microscopic theory for multipole state of
heavy lanthanide systems.

In this Letter, first we define multipole operators of $f$ electrons.
Then, we analyze multipole susceptibility of a seven-orbital Anderson model
by using a numerical renormalization group technique.
It is found that quadrupole moments appear even for Ln=Gd and
an exotic phase governed by octupole moment is suggested for Ln=Ho.
We find no significant multipole moments at low temperatures
for Ln=Tb and Tm, while dipole and higher-order multipole moments
are dominant for Ln=Dy, Er, and Yb.
We discuss these multipole states in comparison with
available experimental results.


The seven-orbital Anderson model is given by
\begin{equation}
 \begin{split}
  H & = \sum_{\mib{k},\sigma}
  \varepsilon_{\mib{k}} c_{\mib{k}\sigma}^{\dag} c_{\mib{k}\sigma}
  +\sum_{\mib{k},\sigma,m}
  (V_{m} c_{\mib{k}\sigma}^{\dag}f_{m\sigma}+{\rm h.c.}) \\
  &+ \sum_{m,m'}\sum_{\sigma,\sigma'} (B_{m,m'}\delta_{\sigma\sigma'}
  + \lambda \zeta_{m,\sigma,m',\sigma'})
  f_{m\sigma}^{\dag}f_{m'\sigma'} \\
  &+ \sum_{m_1 \sim m_4}\sum_{\sigma,\sigma'}
  I_{m_1m_2,m_3m_4}
  f_{m_1\sigma}^{\dag}f_{m_2\sigma'}^{\dag}
  f_{m_3\sigma'}f_{m_4\sigma},
 \end{split}
\end{equation}
where $\varepsilon_{\mib{k}}$ denotes conduction electron dispersion,
$c_{\mib{k}\sigma}$ indicates the annihilation operator for conduction
electron with momentum $\mib{k}$ and spin $\sigma$,
$\sigma$=$+1$ ($-1$) for up (down) spin,
$f_{m\sigma}$ is the annihilation operator for $f$ electron with
spin $\sigma$ and $z$-component $m$ of angular momentum $\ell$=3,
$V_{m}$ is the hybridization between conduction and $f$ electrons,
$B_{m,m'}$ is a crystalline electric field (CEF) potential for
$f$ electrons, $\delta_{\sigma\sigma'}$ is the Kronecker's delta,
$\lambda$ is the spin-orbit coupling,
$\zeta_{m,\pm 1,m,\pm 1}$=$\pm m/2$,
$\zeta_{m \pm 1,\mp 1,m, \pm 1}$=$\sqrt{12-m(m \pm 1)}/2$,
and zero for the other cases.
The Coulomb integral $I$ is expressed by the combination of
Slater-Condon parameters, $F^0$, $F^2$, $F^4$, and $F^6$.
\cite{Slater}

For filled skutterudites, the main conduction band is given by
$a_{\rm u}$, constructed from $p$-orbitals of pnictogen.\cite{Harima}
Note that the hybridization occurs between the states
with the same symmetry.
Since the $a_{\rm u}$ conduction band has xyz symmetry,
we set $V_2$=$-V_{-2}$=$V$ and zero for other $m$.
For the $T_{\rm h}$ point group, $B_{m,m'}$ is given by three
CEF parameters, $B_4^0$, $B_6^0$, and $B_6^2$,\cite{Takegahara}
expressed as $B_4^0$=$Wx/15$, $B_6^0$=$W(1-|x|)/180$,
and $B_6^2$=$Wy/24$.\cite{LLW,Hutchings}
Here $x$ and $y$ specify the CEF scheme,
while $W$ determines the energy scale of the CEF potential.
In order to adjust the local $f$-electron number $n$,
we appropriately change the chemical potential.


Let us set the parameters of the model.
Details of the determination of the parameters
will be discussed elsewhere.
Concerning Slater-Condon parameters, we set $F^0$=10 eV by hand.
Others are determined so as to reproduce excitation spectra of
Pr$^{3+}$ ion.\cite{Pr3+}
Here we show only the results:
$F^2$=8.75 eV, $F^4$=6.60 eV, and $F^6$=4.44 eV.
We use these values for all lanthanides.
On the other hand, we use experimental value of $\lambda$
for each lanthanide such as $\lambda$=0.180 eV (Gd),  0.212 eV (Tb),
0.240 eV (Dy), 0.265 eV (Ho), 0.295 eV (Er), 0.326 eV (Tm),
and 0.356 eV (Yb).\cite{spin-orbit}
The hybridization is fixed as $V$=0.05 eV and a half of the bandwidth
of $a_{\rm u}$ conduction band is set as 1 eV.
Concerning CEF parameters, we set $W$=$-0.4$ meV, $y$=0.3, and $x$=0.3
so as to reproduce quasi-quartet CEF scheme of PrOs$_4$Sb$_{12}$.
\cite{Kohgi,Kuwahara,Goremychkin}


Here we provide a comment on the local CEF state.
Due to the lack of space, we do not show the CEF energy schemes
for $n$=7$\sim$13, but they are almost the same as those obtained
in Ref.~\citen{Hotta3},
even though we have used different parameters there.
Namely, the CEF energy schemes are not changed drastically,
as long as we use realistic Coulomb interaction and
spin-orbit coupling.
As for the CEF ground state, except for $n$=10,
it is not easily converted when we slightly change the value of
$x$ around at $x$=0.3.
Thus, we set $x$=0.3 as a typical value
for $n$=7$\sim$9 and 11$\sim$13.
For $n$=10, it has been found that $\Gamma_{23}^+$ doublet and
$\Gamma_4^+$ triplet states are almost degenerate
in the wide range of the values of $x$.\cite{Hotta3}
Namely, it is necessary to discuss carefully the case of $n$=10.


Now we define the multipole operator ${\hat X}$ for $f$ electron.
In general, ${\hat X}$ is expressed as
\begin{equation}
  \label{multi}
  {\hat X}=\sum_{k,\gamma}
   p^{(k)}_{\gamma}{\hat T}^{(k)}_{\gamma},
\end{equation} 
where $k$ is a rank of multipole,
$\gamma$ is a label to express $O_{\rm h}$ irreducible representation,
and ${\hat T}^{(k)}_{\gamma}$ is cubic tensor operator,
given by ${\hat T}^{(k)}_{\gamma}$=
$\sum_q G^{(k)}_{\gamma,q}{\hat T}^{(k)}_{q}$.
Here an integer $q$ runs between $-k$ and $k$,
${\hat T}^{(k)}_q$ is spherical tensor operator,
and $G^{(k)}_{\gamma,q}$ is the transformation matrix
between spherical and cubic harmonics.
The coefficient $p^{(k)}_{\gamma}$ will be discussed later.


In this paper, we define multipole as spin-orbital density
in the form of one-body operator from the viewpoint of
multipole expansion of electron density in electromagnetism.
Then, the spherical tensor operator ${\hat T}^{(k)}_q$ for $f$ electron
is given in the second-quantized form as
\begin{equation}
  {\hat T}^{(k)}_q = \sum_{m\sigma,m'\sigma'}
  T^{(k,q)}_{m\sigma,m'\sigma'}f^{\dag}_{m\sigma}f_{m'\sigma'},
\end{equation}
where $T^{(k,q)}_{m\sigma,m\sigma'}$ can be calculated
by using the Wigner-Eckart theorem as \cite{Inui}
\begin{equation}
 \label{Tkq}
 \begin{split}
  T^{(k,q)}_{m\sigma,m'\sigma'}
  &= \sum_{j,\mu,\mu'}
  \frac{\langle j || T^{(k)} || j \rangle}{\sqrt{2j+1}}
  \langle j \mu | j \mu' k q \rangle \\
  &\times 
  \langle j \mu | \ell m s \frac{\sigma}{2} \rangle
  \langle j \mu' | \ell m' s \frac{\sigma'}{2} \rangle.
 \end{split}
\end{equation}
Here $\ell$=3, $s$=1/2, $j$=$\ell$$\pm$$s$,
$\mu$ is the $z$-component of $j$,
$\langle j \mu | j' \mu' j'' \mu'' \rangle$ denotes
the Clebsch-Gordan coefficient,
and $\langle j || T^{(k)} || j \rangle$ is
the reduced matrix element for spherical tensor operator,
given by
$\langle j || T^{(k)} || j \rangle$=
$\sqrt{(2j+k+1)!/(2j-k)!}/2^k$.\cite{Inui}
Note that $k$$\le$$2j$ and the highest rank of $f$-electron
multipole is 7 in the present definition.\cite{note}

We note that it is not necessary to take double summations
concerning $j$ in eq.~(\ref{Tkq}),
since the matrix representation of total angular momentum
$\mib{J}$ is block-diagonalized in the $(j,\mu)$-basis.
We have checked that the same results as eq.~(\ref{Tkq})
are obtained when we calculate higher-order multipole operators
by following the symmetrized expression of multiple products of
$\mib{J}$,\cite{Inui,Shiina-multi1,Shiina-multi2}
given in the $(m,\sigma)$-basis as
$\mib{J}_{m\sigma,m'\sigma'}$=
$\delta_{\sigma\sigma'} \mib{L}_{mm'}$+
$\mib{S}_{\sigma\sigma'}\delta_{mm'}$.
Here $\mib{L}$ denotes angular momentum operator for $\ell$=3
and $\mib{S}$ indicates spin operator,
given by $\mib{S}$=$\mib{\sigma}/2$ with Pauli matrix $\mib{\sigma}$.


Let us now determine the coefficient $p^{(k)}_{\gamma}$.
In order to discuss the multipole state, it is necessary to evaluate
the multipole susceptibility in the linear response theory.\cite{Hotta4}
However, multipoles belonging to the same symmetry are mixed in general,
even if the rank is different.
In addition, multipoles are also mixed due to the CEF effect.
Thus, we determine $p^{(k)}_{\gamma}$ by the normalized eigenstate
of susceptibility matrix
\begin{equation}
 \begin{split}
  \chi_{k\gamma,k'\gamma'} \!
  = & \frac{1}{Z}
  \sum_{i,j} \frac{e^{-E_i/T}-e^{-E_j/T}}{E_j-E_i}
  \langle i | [{\hat T}^{(k)}_{\gamma}-\rho^{(k)}_{\gamma}] | j \rangle
  \\ & \times 
  \langle j | [{\hat T}^{(k')}_{\gamma'}- \rho^{(k')}_{\gamma'}]| i \rangle,
 \end{split}
\end{equation}
where $E_i$ is the eigenenergy for the $i$-th eigenstate
$|i\rangle$ of $H$, $T$ is a temperature,
$\rho^{(k)}_{\gamma}$=$\sum_i e^{-E_i/T}
\langle i |{\hat T}^{(k)}_{\gamma}| i \rangle/Z$,
and $Z$ is the partition function given by
$Z$=$\sum_i e^{-E_i/T}$.
Note that the multipole susceptibility is given by
the eigenvalue of the susceptibility matrix.


In order to evaluate the susceptibility matrix, here we employ
a numerical renormalization group (NRG) method,\cite{NRG}
in which momentum space is logarithmically
discretized to include efficiently the conduction electrons
near the Fermi energy.
In actual calculations, we introduce a cut-off $\Lambda$ for
the logarithmic discretization of the conduction band.
Due to the limitation of computer resources,
we keep $M$ low-energy states.
In this paper, we set $\Lambda$=5 and $M$=4500.
Note that the temperature $T$ is defined as
$T$=$\Lambda^{-(N-1)/2}$ in the NRG calculation,
where $N$ is the number of the renormalization step.


Here five comments are in order.
(i) In this paper, we call higher multipoles by following
the numerical terms of IUPAC,\cite{IUPAC}
i.e., 16=hexadeca, 32=dotriaconta, 64=tetrahexaconta,
and 128=octacosahecta.
(ii) We express the irreducible representation of the CEF state
by Bethe notation in this paper, but for multipoles,
we use short-hand notations by the combination of the number
of irreducible representation and the parity of time reversal
symmetry, g for gerade and u for ungerade.
(iii) For $T_{\rm h}$ symmetry, $\Gamma_1$ and $\Gamma_2$ of
$O_{\rm h}$ are mixed.
We remark that $\Gamma_4$ and $\Gamma_5$ of $O_{\rm h}$ are
also mixed.\cite{Takegahara}
Thus, we obtain six independent multipole components as
1g+2g, 2u, 3g, 3u, 4g+5g, and 4u+5u for filled skutterudites.
Note that 1u does not appear within rank 7.
(iv) We normalize each multipole operator so as to satisfy
the orthonormal condition
Tr$\{ {\hat T}^{(k)}_{\gamma}{\hat T}^{(k')}_{\gamma'} \}$=
$\delta_{kk'}\delta_{\gamma\gamma'}$,\cite{Kubo}
when we express the multipole moment as eq.~(\ref{multi}).
(v) The susceptibility for 4u multipole moment does $not$ mean
magnetic susceptibility, which is evaluated by the response of
magnetic moment $\mib{L}$+$2\mib{S}$, i.e.,
$\mib{J}$+$\mib{S}$.\cite{Hotta2,Hotta3}


\begin{figure}[t]
\centering
\includegraphics[width=7.0truecm]{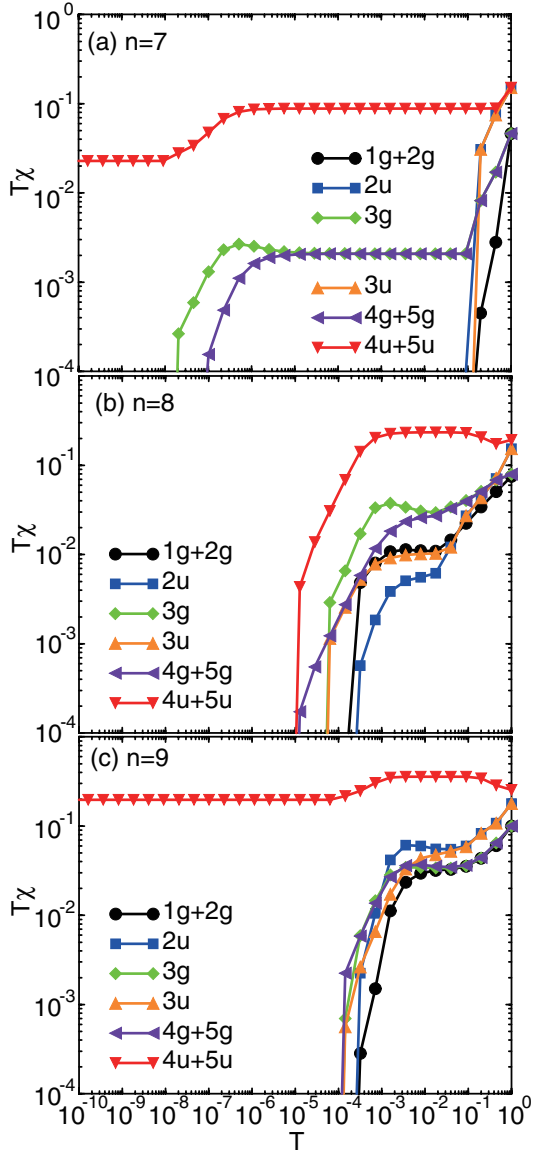}
\caption{(Color online)
$T\chi$ for (a) $n$=7, (b) $n$=8, and (c) $n$=9.
}
\end{figure}

Now we discuss the numerical results.
In Fig.~1(a), we show the results for $n$=7 (Gd$^{3+}$ ion).
Note that we depict only the largest eigenvalue state of
each symmetry group.
The CEF ground state is well described by
$L$=0 and $J$=$S$=7/2, but it is almost independent of $x$,
since the CEF potentials for $L$=0 provide only the energy shift.
The dominant multipole component is 4u dipole and
the contributions from higher multipoles are negligibly small.
We find that the secondary components are 3g and 5g quadrupoles.
Around at $T$=$10^{-7}$, there occurs partial screening
due to conduction electrons and at low temperatures,
spin $S$=$1/2$ remains.\cite{Hotta2,Hotta3}
In such a region, quadrupole components disappear.

In the $LS$ coupling scheme, the $f^7$ state is
uniquely specified by $J$=$S$=7/2 and $L$=0.
Thus, quadrupole does not seem to appear at the first glance.
However, we should note that actual situation is always
deviated from the $LS$ coupling scheme.\cite{Hotta6}
Some finite contribution of the $j$-$j$ coupling scheme
is included in the ground state.
If we consider very large $\lambda$, first we fully occupy $j$=5/2
sextet and then, we accommodate one $f$ electron in $j$=7/2 octet.
Thus, this state can be multipole-active.
Quite recently, it has been observed in GdRu$_4$P$_{12}$
that $^{101}$Ru NQR frequency exhibits temperature dependence
below a N\'eel temperature $T_{\rm N}$=22K.\cite{Kohori}
This may be interpreted as the effect of quadrupole
due to the deviation from the $LS$ coupling scheme.

In Fig.~1(b), we show the results for $n$=8 (Tb$^{3+}$ ion).
At $x$=0.3, the CEF ground state is $\Gamma_1^+$ singlet
and the excited state is $\Gamma_4^+$ triplet with excitation energy
of 1 meV.
For $T$$>$$10^{-4}$, the dominant component is 4u dipole,
while for $T$$<$$10^{-4}$, we find no significant local moments.
Note that we obtain finite values for $\rho^{(4)}_{\rm 1g}$,
$\rho^{(6)}_{\rm 1g}$, and $\rho^{(6)}_{\rm 2g}$
due to the effect of CEF potentials.
Thus, we cannot exclude a possibility of antiferro 1g+2g
ordering.\cite{Takimoto,Sakai}
For TbRu$_4$P$_{12}$, $T_{\rm N}$ is found to be 20K,
but another anomaly is observed at $T_1$=10K.\cite{Sekine}
It may be interesting to discuss $T_1$ in the context of
antiferro hexadecapole and tetrahexacontapole order.

In Fig.~1(c), the results for $n$=9 (Dy$^{3+}$ ion) are depicted.
In this case, the CEF ground state is $\Gamma_5^-$ doublet and
the excited state is $\Gamma_{67}^-$ quartet with excitation energy
of 2.7 meV.
The dominant multipole is always the mixture of 4u and 5u.
The main component is dipole, but we find significant
contribution from octacosahectapole.

\begin{figure}[t]
\centering
\includegraphics[width=7.0truecm]{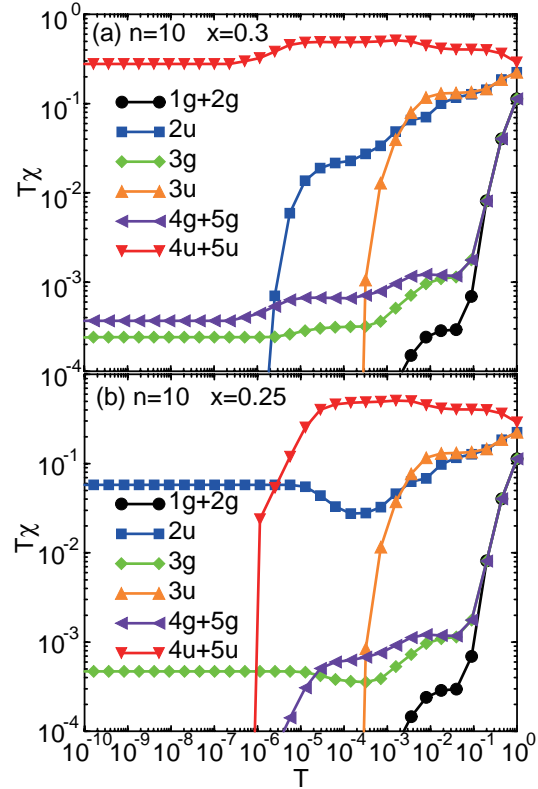}
\caption{(Color online)
$T\chi$ for $n$=10 with (a) $x$=0.3 and (b) $x$=0.25.
}
\end{figure}

Next we focus on the case of $n$=10 (Ho$^{3+}$ ion).
At $x$=0.3, the CEF ground state is $\Gamma_4^+$ triplet,
but the first excited state is $\Gamma_{23}^+$ doublet
with very small excitation energy such as $10^{-5}$ eV.
Thus, Ho-based filled skutterudite is considered to be
in the quasi-quintet situation.\cite{Hotta3,Yoshizawa2}
In Fig.~2(a), we show the results for $n$=10 and $x$=0.3.
We observe several kinds of multipoles, but the dominant one
is always given by the mixture of 4u and 5u
from dipole, octupole, dotriacontapole, and octacosahectapole.

As mentioned above, the CEF ground state is fragile for $n$=10.
If we slightly decrease $x$, the ground state is easily changed.
Then, we evaluate multipole susceptibility for $x$=0.25 with
$\Gamma_{23}^+$ doublet ground state, as shown in Fig.~2(b).
The multipole states at high temperatures are similar to
those in Fig.~2(a), but at low temperatures,
we find 2u multipole state, expressed as
$p^{(3)}_{\rm 2u}$=$0.955$ and $p^{(7)}_{\rm 2u}$=$0.297$.
Namely, the main component is 2u octupole, but there is
contribution of 2u octacosahectapole.
From the elastic constant measurement for HoFe$_4$P$_{12}$,
some anomalous features have been discussed.\cite{Yoshizawa2}
It may be interesting to reexamine experimental results
from the viewpoint of 2u octupole state.

\begin{figure}[t]
\centering
\includegraphics[width=7.0truecm]{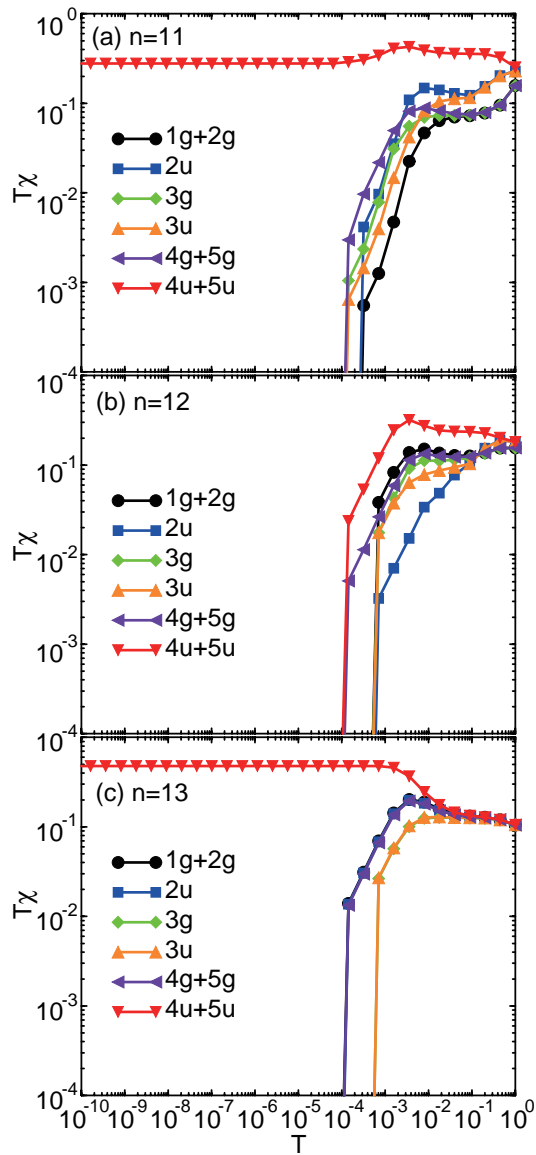}
\caption{(Color online)
$T\chi$ for (a) $n$=11, (b) $n$=12, and (c) $n$=13.
}
\end{figure}

Let us move onto the cases of $n$=11, 12, and 13.
In Fig.~3(a), we show the results for $n$=11 (Er$^{3+}$ ion).
For $n$=11, the CEF ground state is $\Gamma_5^-$ doublet and
the excited state is $\Gamma_{67}^-$ quartet.
The excitation energy is 5.6 meV.
We find that the dominant multipole state is given by 4u+5u,
but main components are 4u dipole and 4u octacosahectapole.
The admixture is almost independent of temperature.

In Fig.~3(b), the results for $n$=12 (Tm$^{3+}$ ion) are depicted.
The CEF ground state at $x$=0.3 is $\Gamma_1^+$ singlet,
while the excited state is $\Gamma_4^+$ triplet
with the excitation energy of 5.9 meV.
In the high-temperature region, several kinds of multipoles appear,
but when we decrease temperature, there appear no significant
multipole moments, as in the case of $n$=8.
Again we mention that a possibility of antiferro 1g+2g ordering
cannot be excluded.

In Fig.~3(c), we depict the results for $n$=13 (Yb$^{3+}$ ion).
The CEF ground and excited states are both $\Gamma_5^-$ doublets
with the excitation energy of 5.1 meV.
When temperature is decreased, there appears the dominant
mixed multipole of 4u and 5u with significant components
of dipole, octupole, dotriacontapole, and octacosahectapole.
Among them, large contributions come from dotriacontapole
and octacosahectapole.


Finally, let us briefly discuss the Kondo effect.
First we note that the so-called underscreening Kondo effect can occur
in the present model, since we include only the single $a_{\rm u}$
conduction band hybridized with some of seven $f$ orbitals.
In fact, we have found the partial screening for $n$=7,
but for other values of $n$, we did not observe the Kondo behavior
in the present temperature range,
since the Kondo temperature becomes much smaller than the CEF splitting.
However, if $V$ is increased, we expect to
find the Kondo behavior for $n$$\ne$7.
When we perform the calculations for $V$=0.5,
we actually observe that the residual 4u moment is eventually screened
even in the present temperature range for $n$$\ne$7.
Note, however, that the Kondo behavior can be found only
for unrealistic large value of $V$ in the present model.
In order to discuss appropriately the Kondo effect in filled skutterudites,
it is highly required to include not only $a_{\rm u}$ but also $e_{\rm u}$
conduction bands.
It is one of future problems.


In summary, we have discussed possible multipole states of
filled skutterudites including heavy lanthanide ion Ln$^{3+}$.
For Ln=Gd, we have found the effect of quadrupole moments
due to the deviation from the $LS$ coupling scheme.
For Ln=Ho, the CEF ground state is quasi-quintet
with $\Gamma_{23}^+$ doublet and $\Gamma_4^{+}$ triplet.
When we slightly change the CEF potential, the exotic state
dominated by 2u octupole moment has been found.
For Ln=Tb and Tm, the CEF ground state is $\Gamma_1^+$ singlet and
we have found no significant multipole moments at low temperatures,
although we cannot exclude a possibility of antiferro 1g+2g ordering.
For Ln=Dy, Er, and Yb, the CEF ground state is $\Gamma_5^-$ doublet,
and the dominant moment is the mixture of 4u and 5u.
We believe that the present results stimulate further study of
multipole phenomena in heavy lanthanide filled skutterudites.


The author thanks K. Kubo for useful discussions on multipoles.
He also thanks H. Harima, Y. Kohori, H. Onishi, C. Sekine,
and M. Yoshizawa for comments.
This work is supported by the Japan Society for the Promotion
of Science and by the Ministry of Education, Culture, Sports,
Science, and Technology of Japan.
The computation in this work has been done using the facilities
of the Supercomputer Center of Institute for Solid State Physics,
University of Tokyo.



\begin{thebibliography}{99}
\bibitem{note}
When we define multipole as tensor operator in the space of
total angular momentum $J$, the highest rank of multipole is $2J$
and there appear multipoles with $k \ge 8$ for the case of $J \ge 4$.
If we need such higher-order multipoles with $k \ge 8$, they are
defined as many-body operators in the second-quantized form.
\end{thebibliography}
\end{document}